\documentclass[journal=ancac3,manuscript=article]{achemso}

\usepackage{chemformula}
\usepackage{siunitx}
\usepackage{gensymb} 
\usepackage[version=4]{mhchem}
\usepackage[T1]{fontenc} 
\usepackage{float}
\usepackage{amssymb}

\usepackage{newtxmath}
\usepackage[scr=rsfso]{mathalfa}

\usepackage{tabularx}
\usepackage{enumitem}
\usepackage{soul}

\usepackage{mathtools}

\DeclarePairedDelimiter\ket{\lvert}{\rangle}
\DeclarePairedDelimiterX\braket[2]{\langle}{\rangle}{#1\,\delimsize\vert\,\mathopen{}#2}

\author{Pavel V. Kolesnichenko}
\affiliation[Heidelberg University]{Physikalisch-Chemisches Institut, Ruprecht-Karls-Universität Heidelberg, 69120, Heidelberg, Germany}
\alsoaffiliation[CAM]{Institute for Molecular Systems, Engineering and Advanced Materials, Ruprecht-Karls-Universität Heidelberg, 69120, Heidelberg, Germany}
\email{pavel.kolesnichenko@alumni.uni-heidelberg.de}

\author{Manuel Hertzog}
\affiliation[Heidelberg University]{Physikalisch-Chemisches Institut, Ruprecht-Karls-Universität Heidelberg, 69120, Heidelberg, Germany}

\author{Felix Hainer}
\affiliation[Heidelberg University]{Physikalisch-Chemisches Institut, Ruprecht-Karls-Universität Heidelberg, 69120, Heidelberg, Germany}
\alsoaffiliation[CAM]{Institute for Molecular Systems, Engineering and Advanced Materials, Ruprecht-Karls-Universität Heidelberg, 69120, Heidelberg, Germany}

\author{Oskar Kefer}
\affiliation[Heidelberg University]{Physikalisch-Chemisches Institut, Ruprecht-Karls-Universität Heidelberg, 69120, Heidelberg, Germany}
\alsoaffiliation[CAM]{Institute for Molecular Systems, Engineering and Advanced Materials, Ruprecht-Karls-Universität Heidelberg, 69120, Heidelberg, Germany}

\author{Jana Zaumseil}
\affiliation[Heidelberg University]{Physikalisch-Chemisches Institut, Ruprecht-Karls-Universität Heidelberg, 69120, Heidelberg, Germany}

\author{Tiago Buckup}
\affiliation[Heidelberg University]{Physikalisch-Chemisches Institut, Ruprecht-Karls-Universität Heidelberg, 69120, Heidelberg, Germany}
\alsoaffiliation[CAM]{Institute for Molecular Systems, Engineering and Advanced Materials, Ruprecht-Karls-Universität Heidelberg, 69120, Heidelberg, Germany}
\email{tiago.buckup@pci.uni-heidelberg.de}

\title{Dark excitons and hot electrons modulate exciton-photon strong coupling in metal-organic optical microcavities}


\begin{document}

\clearpage

\begin{abstract}

Polaritons, formed as a result of strong hybridization of matter with light, are promising for important applications including organic solar cells, optical logic gates, and qubits. Owing to large binding energies of Frenkel excitons (matter), strong matter-light coupling phenomena are possible at room temperature, high exciton densities, and even with low-quality-factor microcavities. In such cases, due to polaritons' high degree of delocalization, simultaneous effects from dark excitons and hot electrons may affect performance of potential devices. Their understanding, therefore, is of paramount importance, but their disentanglement in optical spectroscopy, however, thus far remained unattainable. Here, we overcome this challenge by careful and systematic analysis of transient polaritonic spectra, supported by analytical models. In doing so, we conclude that dark excitons affect the strength of exciton-photon coupling and manifest themselves as Fano-like polaritonic gain-loss spectra. Free electrons add additional loss component to and imprint a two-temperature dynamics on the polaritonic response. The developed general methodology can be applied to a variety of other microcavity structures. Our findings are significant for distinguishing polaritons and other excitations in studies of polariton-electron and plasmon-electron coupling phenomena as well as photonic control over photophysical and photochemical processes.

\end{abstract}

\section{Introduction}

Polaritons are hybrid quasi-particles formed as a result of strong interactions of light and matter, inheriting properties of both. They hold promise for numerous technological advancements such as terahertz sources and detectors\cite{Lee2020}, low-threshold lasers\cite{Deng2003}, all-optical transistors\cite{Ballarini2013}, refractive-index sensors\cite{Xu2019}, organic solar cells\cite{Nikolis2019}, spatial light modulators\cite{JingQuan2008}, and qubits\cite{Ghosh2020}. On a more fundamental side, they have been used in studies of Bose-Einstein condensates (BECs)\cite{Kasprzak2006}, superfluids\cite{Amo2009}, and acoustic black holes\cite{Nguyen2015}, to name just a few.

The strong coupling of light to matter and related phenomena have commonly been studied using Wannier-Mott excitons in the  active (semiconductor) layer of optical microcavities (OMs)\cite{Luo2023}. Recently, organic matter has gained popularity because it can provide a high density of Frenkel excitons with large binding energies, allowing investigations of polaritonic phenomena at room temperature and even with low-quality-factor (low-Q) microcavities\cite{Hobson2002,Pradeesh2009,Schwartz2012,Zhong2016,DelPo2020,Renken2021,Hirai2023}. This, in turn, has allowed the observation of polariton-enhanced phenomena such as superabsorption for energy storage\cite{Quach2022}, enhanced photoluminescence for lasing applications\cite{Ballarini2014}, and, for optoelectronic applications, enhanced energy-transfer through heterjunctions\cite{Wang2021}, enhanced singlet fission (SF)\cite{MartinezMartinez2018} and microsecond-long polariton lifetime\cite{Polak2020}. The scenarii of polariton-assisted enhancement of useful effects, however, may not always occur\cite{Du2018,MartinezMartinez2018,Liu2020,Gu2021}. For instance, mesoscale energy transfer may not result from the strong coupling of surface plasmons to chromophores\cite{Du2018}. Additionally, exciton fission efficiency during SF can decrease either through the photonic leakage of polaritonic states\cite{MartinezMartinez2018}, or, when mediated by conical intersection, due to poorer Franck-Condon overlap emerging in the hybrid system\cite{Gu2021}. This indicates that, although polaritons have been reasonably well understood on a fundamental level, their remarkable physical properties in realistic applications may be affected by multiple other processes occurring in parallel. Understanding these processes, their effects on polaritons, their disentanglement, and potential harvest are therefore of paramount importance.

Metal-based OMs are among the most recently popular physical systems for the realization of strong coupling phenomena\cite{Hobson2002,Pradeesh2009,Schwartz2012,Zhong2016,DelPo2020,Renken2021,Hirai2023,Ballarini2014,Wang2021,Polak2020,Liu2020,Lundt2016,BarraBurillo2021}. 
In such systems, understanding the effects of dark excitons (organic layer) and free electron gas (metal cladding) is of great interest. This is not only because they can affect the performance of novel devices\cite{Li2016,Michail2024}, but also due to their potential role in studies of interactions between electron gases and polaritons\cite{Plyashechnik2023}, BECs\cite{Sun2021a}, or 
conduction electrons\cite{Ulstrup2024}, rendering metal-based OMs a promising platform for studying such interactions. Optical spectroscopy remains the main approach to probe strong-coupling physics in OMs; however, disentangling the excitations in different microcavity layers has remained a challenge, and their manifestation in the nonlinear optical response of polaritons is still unclear.

In particular, recent studies of SF in metal-based OMs indicated that, instead of enhancing the yield of triplet excitons, polaritons decayed to the manifold of dark states much faster than even cavities' natural photonic leakage\cite{Liu2020,Liu2021,Climent2022}. The seemingly contradictory long-lasting nonlinear optical response of polaritons pointed to the dominant effects of dark excitons. Later experiments showed that even when selectively exciting the free electron gas using infrared photons, polaritonic signatures were still detectable near the main exciton resonance in the visible\cite{Liu2021}. In these experiments, there was no pump-induced population of polaritonic states; thus, the observed signals represented strong-coupling conditions rather than long-lived polaritons\cite{Renken2021}. We note that, in this case, the Rabi energy quantifying strong coupling is expected to change with changes in photonic environment, which, however, was not observed and remains to be established. Overall, these case-studies support that the polaritons' strong optical response generally contains mixed dynamics of both dark excitons and the free electron gas, but their disentanglement, as mentioned, has not yet been achieved in optical experiments.

Here, we disentangle the main contributions to the transient optical response of all-metal OMs using TIPS-PEN in the active layer under strong coupling conditions. Supported by theoretical analysis, polaritons were found to be strongly delocalized over both organic and metal layers, inheriting information on the dynamics of dark excitons and hot electrons, respectively, modulating photonic environment and exciton-photon coupling. The coupling strength, quantified through the Rabi energy, was found to be strongly correlated with the dynamics of dark-exciton fission in the organic layer. Such excitons were shown to contribute additional phase offsets to polaritonic resonances responsible for changes in the Rabi energy. The free electron gas, on the other hand, was assigned a role of contributing additional loss signal near the energies of polaritonic modes, evolving in accordance with a two-temperature model and further shaping the nonlinear optical response. Overall, excitations across different cavity layers give rise to Fano-like optical gain-loss spectra at polariton energies. Its systematic interpretation has not been described previously, and our general approach paves the way toward a better understanding of excited-state phenomena in various other microcavity structures.

\section{Results and discussion}

Two cavities (thin and thick, see Supplementary material, Section S1, for more details) with different thicknesses of the organic layer have been considered in this work. Unless stated otherwise, the thin cavity is described further in the text, while the results on the thick cavity are given in the Supplementary material. 

\subsection{Strong coupling and polariton delocalization}

The angle-resolved steady-state reflectance spectrum of polaritons is shown in Fig.~\ref{fig:fig1}a, with hybridized modes (solid lines) indicated (see Supplementary material, Section~S2, for more details). These modes reflect strong coupling between excitons ($X$ and $X'$) and photons (cavity mode \textit{C}) giving rise to quasi-particles with intermediate effective masses (band curvatures) referred to as exciton-polaritons or, simply, polaritons. The two lowest-energy polaritons, denoted as upper (UP) and lower (LP) polaritons, are separated by $\sim$110~meV, indicating strong coupling. Since the bare excitonic bands are not discernible, they are referred to here as the bands of dark excitons, \textit{i.e.}, excitons that have not coupled strongly to the cavity mode. 

\begin{figure}[ht!]
\centering
\includegraphics[width=1\linewidth,height=\textheight,keepaspectratio]{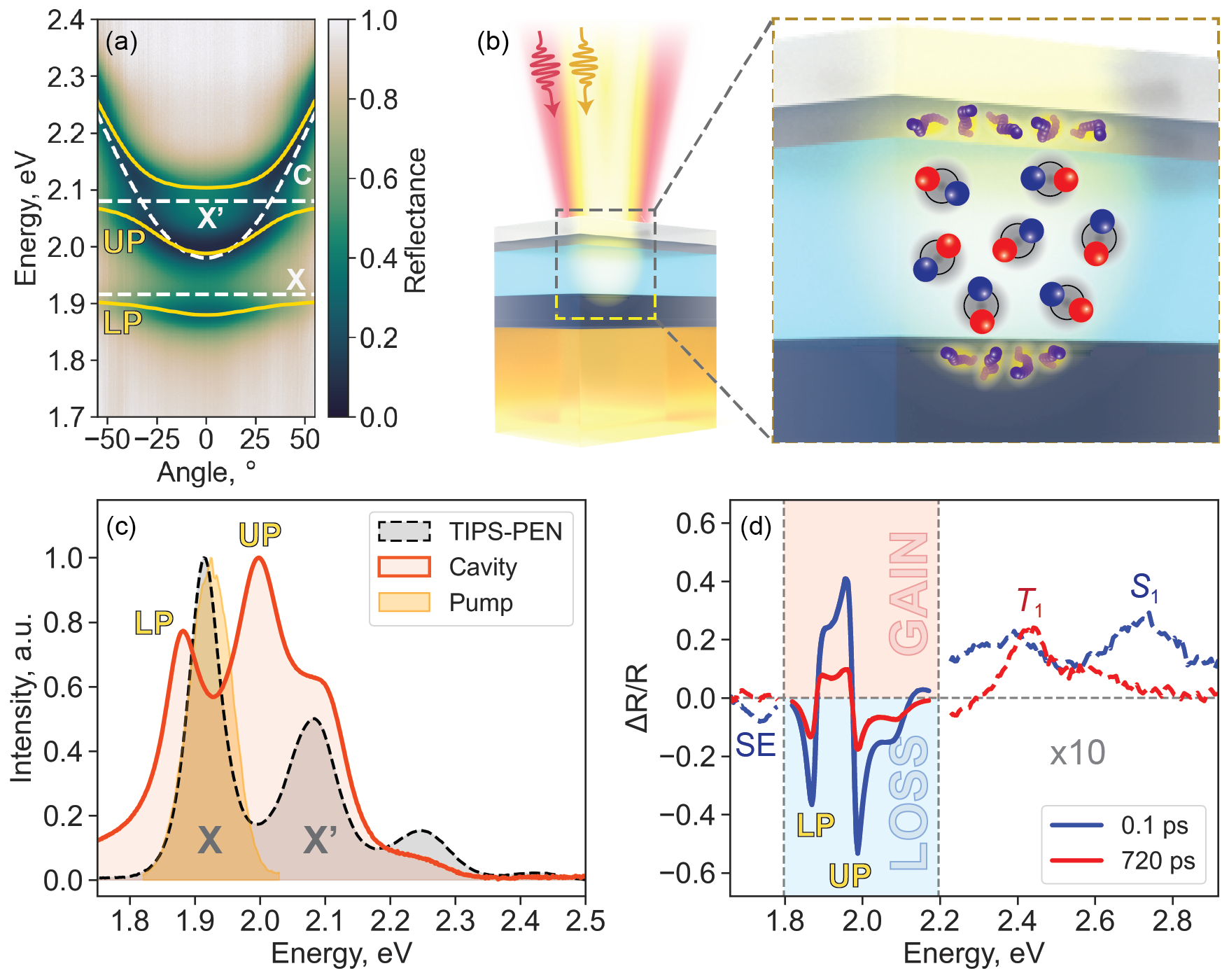}
\caption{Steady-state and transient spectra of thin microcavity. (a) Normalized angle-resolved steady-state reflectance spectrum, with the upper and lower polaritons (UP and LP), dark excitons ($X$ and $X'$), and cavity mode (\textit{C}) indicated. Dashed lines represent uncoupled excitonic (horizontal lines) and photonic (parabola) states; solid yellow lines represent hybridized (polaritonic) states. (b) Transient reflectance experiment. The inset (right) shows dark excitons in the organic layer and hot electrons in metal films "dressed" by polaritonic field.  (c) Steady-state absorption spectrum of the cavity and bare TIPS-PEN (dashed and shaded in grey), and pump spectrum (shaded in yellow). (d) Transient reflectance spectrum at short (0.1~ps, blue) and long (720~ps, red) pump-probe delays. The weak molecular response (dashed) formed from $S_1$ ESA (excited-state absorption) and SE (stimulated emission) at short delays and $T_1$ ESA (red) at long delays is scaled by a factor of 10.}
\label{fig:fig1}
\end{figure}

More formally, the two significant polaritons (UP, $\ket{\psi_{+}}$, and LP, $\ket{\psi_{-}}$) are the result of the hybridization of uncoupled states of exciton ($\ket{\psi_{X}}$ and $\ket{\psi_{X'}}$) and photon ($\ket{\psi_{C}}$), \textit{i.e.},
\begin{equation}
\begin{aligned}
\ket{\psi_{\pm}} = \sqrt{\alpha_{\pm}}\ket{\psi_{X}}+\sqrt{\beta_{\pm}}\ket{\psi_{X'}}+\sqrt{\gamma_{\pm}}\ket{\psi_{C}}.
\end{aligned}
\label{eqn:hopfield}
\end{equation}
In Eq.~(\ref{eqn:hopfield}), other, less intense, vibronic states participating in hybridization are omitted due to their negligible contribution; $\alpha_{\pm},\beta_{\pm},\gamma_{\pm}$ are Hopfield coefficients defining the curvatures of polaritonic bands (Fig.~\ref{fig:fig1}a) by indicating the excitonic ($\alpha_{\pm},\beta_{\pm}$) and photonic ($\gamma_{\pm}$) contribution to each of them. Thus, in the case of the thin cavity, UP inherits a larger photonic fraction compared to LP, whereas in the case of the thick cavity, it is the LP band that is more photonic than UP (see Supplementary material, Section~S3, for estimations of Hopfield coefficients). In this context, it is worth mentioning that there is a third polaritonic band present at $\sim$2.1~eV, which is more excitonic in nature (at 0$\degree$), and therefore not explicitly considered further. 

Our \textit{ab initio} optical simulations (Supplementary material, Section~S4) showed that the more photonic polaritons result in a larger absorption strength (Fig.~\ref{fig:fig1}a,c) due to the cavity-facilitated local field enhancement effects\cite{Kolesnichenko} (a similar trend was observed previously for amorphous rubrene\cite{Takahashi2019}).  Importantly, both polaritons are found not only delocalized over the organic layer but also extended into the metal cladding, where a more photonic polariton also features stronger absorption. Such strong polariton delocalization renders polaritons susceptible to carrier dynamics in both organic and metal films (Fig.~\ref{fig:fig1}b). In particular, it is anticipated that polaritons launch correlated dynamics of both dark excitons and hot electrons, which in turn dynamically modifies exciton-photon coupling. Therefore, the dynamics of the coupling strength, inheriting information on the dynamics of both types of carriers, is considered next.

\subsection{Pump-induced dynamics of exciton-photon coupling}

To gain insights into exciton-photon coupling dynamics, transient reflectance (TR) experiments were carried out (Fig.~\ref{fig:fig1}b and Supplementary material, Section~S5). Due to the small incidence angles of the pump and probe beams, it is the regions close to the bands' extrema in Fig.~\ref{fig:fig1}a that were predominantly probed. The pump was spectrally tailored to overlap the $X$-resonance of uncoupled excitons (Fig.~\ref{fig:fig1}c), thus exciting polaritons, hot electrons, and the dark-states reservoir. The differential reflectance (DR) spectrum of the cavity (Fig.~\ref{fig:fig1}d) features a long-lasting strong dispersive response at the energies of polaritonic bands, which has also been observed previously\cite{Schwartz2012,Xiang2019a,Liu2020, Liu2021}, but the nature of signal gain and loss has not been fully understood. This response greatly dominates over the weak molecular response formed from stimulated emission (SE) in the lower-energy region (1.7--1.8~eV) and $S_1/T_1$ excited-state absorption (ESA) in the higher-energy region (2.3--2.9~eV). The latter corresponds to the optical messengers of SF in TIPS-PEN\cite{Kolesnichenko}. One of the striking differences between thin and thick cavities (Supplementary material, Section~S1) is the amplitude of the reflectance loss (negative) near UP and LP energies at early pump-probe delays (<~10~ps). At longer pump-probe delays (>~10~ps), these dramatic differences between the two samples vanish.

It is important to point out that the lifetime of such response greatly exceeds the expected lifetimes of both UP and LP. Indeed, the estimated cavity-mode lifetime in this study is $\tau_C\sim{9}$~fs (cavities' Q-factor is $\sim{45}$), so the polaritons' decay rate $\tau^{-1}_{\pm}=(\alpha_{\pm}+\beta_{\pm})\tau^{-1}_X+\gamma_{\pm}\tau^{-1}_C \gtrsim 30^{-1}$~fs$^{-1}$ falls far below the time-resolution of our experiments ($\sim$100~fs). Therefore, polaritons are formed predominantly within the duration of the pump and probe pulses and decay into the dark-states manifold within $\sim$30~fs following ever-vanishing pulse tails. The pump pulses therefore effectively generate long-lived excitons in the organic layer and hot electrons in metal films; the probe pulses generate short-lived polaritons again, and it is these probe-induced polaritons that experience changing strong coupling conditions (due to excitations in the organic and metal layers). The reflectance difference between pump-perturbed ($R_{\textrm{ON}}$) and unperturbed ($R_{\textrm{OFF}}$) cavities results in dispersive gain-loss spectra.

Preliminary lineshape analysis (Supplementary material, Section~S6) and fittings with Fano formulae\cite{Fano1961,Limonov2017,Caselli2018} (Supplementary material, Section~S7) showed that the transient spectra are inseparable in time and energy, indicating additional interference effects at play. The fittings also indicated the presence of a weak signal from dark excitons at $\sim$1.92~eV within the overwhelming polaritonic response. Altogether, the evolution of the extracted parameters resembled the course of SF known to occur in TIPS-PEN, suggesting that it is at least the dynamics of dark singlet and triplet excitons that modulated the polaritonic response simultaneously in energy and time. 

Notably, similar Fano-like lineshapes have been observed in various other systems where interference of either electromagnetic waves or probabilistic wavefunctions occurred.\cite{Lin2012,Liao2017,Levi2016,Barontini2022,Salikhov2016,Mun2018,Qi2014,Zhang2022,Owens2010} Most relevant to this work are induced transmission filters (ITFs)\cite{Owens2010}, where a strong dispersive DR response resulted from coupling of a 30-nm silver film to a Lorentz-like oscillator, revealing intrinsic strong nonlinearities in the metal layer. Such nonlinear effects of semi-transparent silver films could also contribute to the measured overwhelming nonlinear response in this work. The extracted time-varying Fano-parameters would then reflect net dynamics arising from contributions from both dark excitons and hot electrons.

Fano formulae describe interband splitting of hybridized bands but not the Rabi energy. Therefore, next, a non-Hermitian formalism is developed, accounting for interference in the differential-reflectance domain and thus describing the observed entangled spectro-temporal behaviour of the strong polaritonic response.

\subsection{Coupling strength modulated by singlet-fission dark excitons}

 A coupled exciton-photon system can be described by an effective non-Hermitian Hamiltonian (see also Supplementary material, Section~S8),
\begin{equation}
\begin{aligned}
\mathbf{H} = \mathbf{E} + \mathbf{V} + \mathbf{\Gamma},
\end{aligned}
\label{eqn:H}
\end{equation}
\begin{equation}
\begin{aligned}
\mathbf{E} = \begin{pmatrix} E_X & 0\\ 0 & E_C \end{pmatrix}, \mathbf{V} = \begin{pmatrix} 0 & V\\ V^* & 0 \end{pmatrix}, \mathbf{\Gamma} = -\frac{i}{2}\begin{pmatrix} \Gamma_X & 0\\ 0 & \Gamma_C \end{pmatrix},
\end{aligned}
\label{eqn:EVG}
\end{equation} 
where $\mathbf{E}$ is the unperturbed Hamiltonian containing the energies ($E_X$, $E_C$) and states ($\ket{\psi_X}$, $\ket{\psi_C}$) of uncoupled exciton and photon; $\mathbf{V}$ describes the coupling between them ($V=\hbar\Omega$ is the coupling strength, $\Omega$ is the Rabi frequency); and $\mathbf{\Gamma}$ describes the spectral widths of exciton ($\Gamma_X$) and photon ($\Gamma_C$). Diagonalization of the Hamiltonian $\mathbf{H}$ (Eq.~(\ref{eqn:H})) yields steady-state eigenfunctions $\ket{\psi_{\pm}}$ and eigenenergies $\mathcal{E}_{\pm}=E_{\pm}-i\Gamma_{\pm}$ of the two hybrid oscillators, where $E_{\pm}$ and $\Gamma_{\pm}$ are their energies and widths, respectively. We note that although there are two vibronic resonances, $X$ and $X'$, contributing to the polaritonic bands, they can be regarded as one effective exciton (see Supplementary material, Section~S9), which greatly simplifies analysis yet still allows for main trends to be extracted.

As demonstrated in the Supplementary material, Section~S7, the measured dispersive lineshapes can be nearly perfectly described with Fano formulae, suggesting they result from interference phenomena. In this scenario, two "net" phase offsets $\phi_{\pm}$ in the two hybrid modes must be considered to accurately describe the combined carrier dynamics in organic and metal films. Since we measured the \textit{changes} of the parameters relative to the initial state (initiated by the pump pulse), we represent the state at any time instance after the pump excitation as a superposition of eigenfunctions of the Hamiltonian $\mathbf{H}$ (Eq.~(\ref{eqn:H})), \textit{i.e.},
\begin{equation}
\begin{aligned}
\ket{\psi} = e^{i\phi_-}\cos{\theta}\ket{\psi_-}+e^{i\phi_+}\sin{\theta}\ket{\psi_+},
\end{aligned}
\label{eqn:initial}
\end{equation}
where $\theta$ controls the weights of contributions from each of the polaritonic states, and $\phi_{\pm}$ are phase offsets relative to the global reference phase $\phi_0=0$ defined by the coupling elements in $\mathbf{V}$.
Then, following the Green-function-based approach\cite{Moiseyev1978,Kukulin1989,Durand2001,Paidarova2003}, transient reflectance can be expressed as
\begin{equation}
\begin{aligned}
\frac{\Delta{R}}{R} = -\frac{A}{\pi}\mathfrak{Im}({G(E)}), \textrm{ with } G = \frac{f_-}{E-\mathcal{E}_{-}} + \frac{f_+}{E-\mathcal{E}_{+}},
\end{aligned}
\label{eqn:dR}
\end{equation}
where $f_{\pm}$ are the generalized differential oscillator strengths of the two polaritonic branches defined through $c$-products between the initial and final states,
\begin{equation}
\begin{aligned}
f_{\pm}=(\psi\lvert\psi_{\pm})(\psi_{\pm}\lvert\psi);
\end{aligned}
\label{eqn:oscstr}
\end{equation}
and $A$ is the scaling factor. We emphasize that since $c$-products (Eq.~(\ref{eqn:oscstr})) are utilized within the non-Hermitian framework, the phase factors of the polaritonic state-vectors are incorporated into the expressions for the differential oscillator strengths. As a result, nontrivial Fano-like resonances naturally emerge.

\begin{figure}[ht!]
\centering
\includegraphics[width=0.9\linewidth,height=\textheight,keepaspectratio]{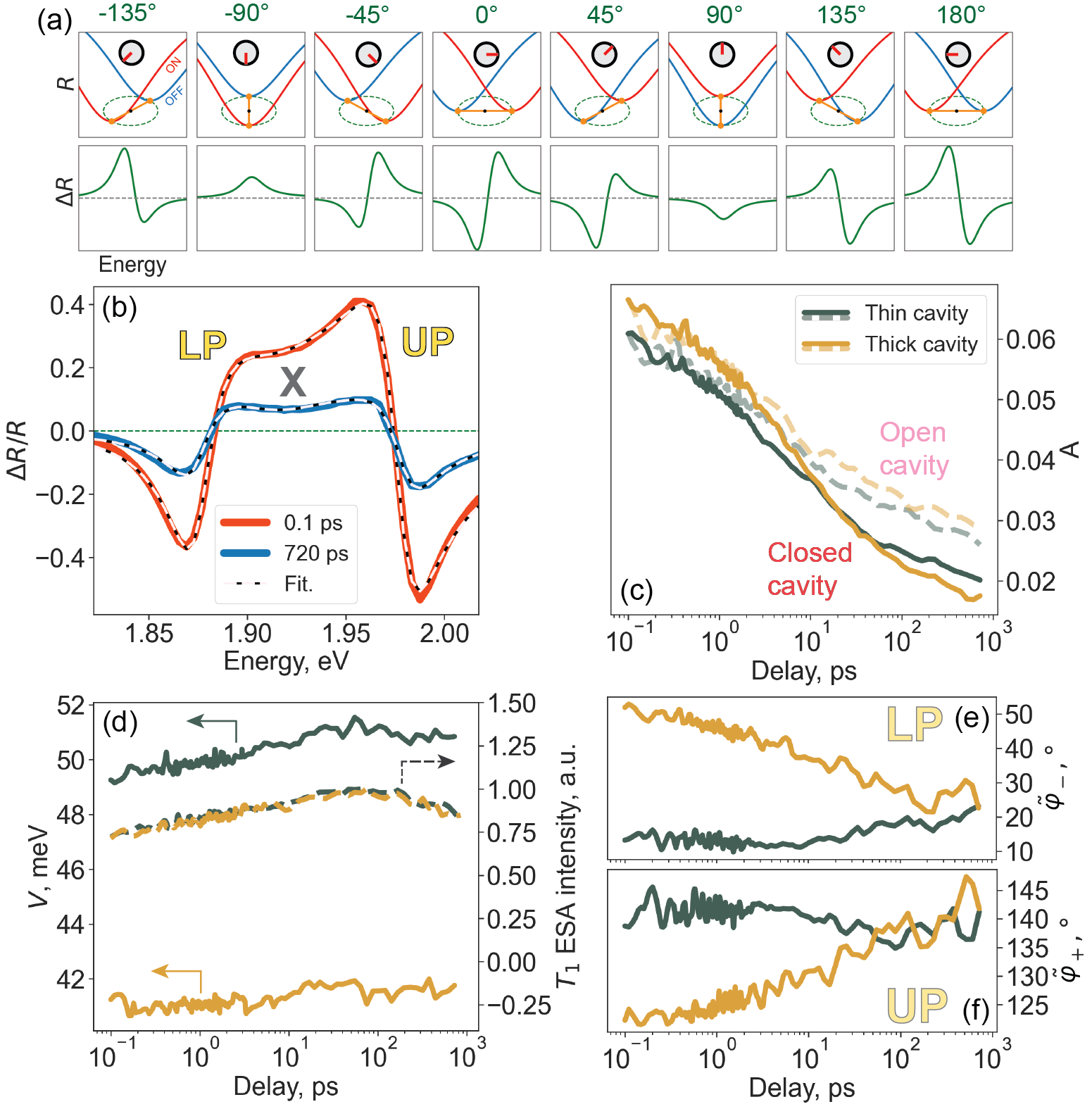}
\caption{Non-Hermitian Hamiltonian description of transient reflectance spectra. (a) A cartoon explaining the physical meaning of the dynamic phases $\phi_{\pm}$ after remapping them onto more intuitive values $\widetilde{\phi}_{\pm}=90\degree+2\phi_{\pm}$ (modulo 180$\degree$). (upper row) "Pump-ON" (red) and "pump-OFF" (blue) reflectance spectra. The inset shows the phase $\widetilde{\phi}$ and the direction of the displacement (red line) of the "pump-ON" spectrum relative to the "pump-OFF" spectrum. (b) Fittings (dashed lines) of short- (red) and long-delay (blue) spectra (solid lines).  (c--f) The scaling amplitude $A$ (c, solid), coupling strengths $V$ (f, solid), and phases $\widetilde{\phi}_{\pm}$ (e,f) retrieved from fittings for thin (dark lines) and thick (yellow lines) cavity. In (c), dashed lines represent the absolute value of the bleach signal from the corresponding open cavities (\textit{i.e.}, cavities without 30-nm metal film). In (f), dashed lines represent $T_1$~ESA in thin and thick cavity.}
\label{fig:fig2}
\end{figure}

Fig.~\ref{fig:fig2}a illustrates the intuition behind the introduced phases $\phi_{\pm}$, which, for this sake, are remapped onto $\widetilde{\phi}_{\pm}=(90\degree+2\phi_{\pm})\mod{180\degree}$. The values of $\widetilde{\phi}_{\pm}$ simply indicate the peak position of the reflectance spectrum $R_\textrm{ON}$ relative to $R_\textrm{OFF}$ in the intensity-energy domain. Thus, $\widetilde{\phi}_{\pm}=90\degree$ ($-90\degree$) correspond to the signal loss (gain), indicating that $R_\textrm{ON}$ is overall "above" ("below") $R_\textrm{OFF}$. The values of  $\widetilde{\phi}_{\pm}=0\degree,180\degree$ correspond to more intense purely dispersive lineshapes. Other values describe dispersive profiles with either a loss or gain component dominating. The $R_\textrm{ON}$ and $R_\textrm{OFF}$ peaks slide along a phase-ellipse around their average location; its shape is ultimately fixed by the optical density of the organic layer, pump fluence, and other experimental conditions. Such qualitative picture is fully consistent with the general cyclic correlation between temporal phase ($\delta{\phi}$) and spectral shift ($\delta{E}$), $\delta{E}\sim\cos{\delta{\phi}}$, for two classical harmonic oscillators\cite{Tolmachev2002}, and that such phase is mappable to the Fano parameter describing gain-loss spectra\cite{Ott2013}. 
Within the developed framework, the coupling strength $V$ reflects only the average value between "pump-OFF" and "pump-ON" conditions. The $R_\textrm{OFF}$ spectrum, however, is time-independent, and therefore the dynamics of the Rabi energy $V$ reflects the pump-induced dynamics encoded in the $R_\textrm{ON}$ spectrum alone.

The fitting of the polaritonic response with the described model is of remarkable quality (Fig.~\ref{fig:fig2}b), given the model's simplicity (see Supplementary material, Sections~S10,S11). Global exponential fitting (Supplementary material, Section~S12) of the dynamics of the extracted parameters revealed four time constants similar to exciton dynamics in bare films and open OMs (\textit{i.e.}, cavities without 30-nm metal film)\cite{Liu2020,Kolesnichenko}, although their values are not the same due to the different photonic environment\cite{Kolesnichenko}.  Fig.~\ref{fig:fig2}c--f show the evolution of important parameters (for thin and thick cavity) describing the decay of the polaritonic response (Fig.~\ref{fig:fig2}c), coupling strength (Fig.~\ref{fig:fig2}d), and phases (Fig.~\ref{fig:fig2}e,f). The amplitude $A$ of the nonlinear response (Fig.~\ref{fig:fig2}c) decays, as mentioned, similarly to exciton bleach in open cavities, which is in turn correlated with the course of SF (where the amount of dark triplet excitons peaks on a sub-100 ps time scale). Importantly, the dynamics of the Rabi energy $V$  (Fig.~\ref{fig:fig2}d) are also strongly correlated with the dynamics of dark triplet excitons generated via SF, which was not observed previously\cite{Liu2020}. 

Remarkably, the phase offsets $\widetilde{\phi}_{\pm}$ (Fig.~\ref{fig:fig2}e,f) for the two samples evolve in opposite directions and converge to the same values for long pump-probe delays, where most of the initial excitations are relaxed and the dynamics are mostly governed by the dark triplet excitons. Their values for UP and LP lie on the opposite sides of the case of a pure bleach-like spectrum ($\widetilde{\phi}_{\pm}=90\degree$ in Fig.~\ref{fig:fig2}a), additionally indicating the presence of a Lorentz-like oscillator (dark exciton) having a resonance located between the two polaritonic modes. Overall, the phases and the direction of their evolution are in full agreement with the intuitive picture in Fig.~\ref{fig:fig2}a, and indicate relaxation of the dispersive response towards a more symmetric lineshape with a dominating loss signal. To gain further insights into retrieved phases and the constitution of the observed gain-loss spectra, we took advantage of \textit{ab initio} electromagnetic simulations based on the (differential) transfer-matrix (TM) method\cite{Pettersson1999,Hecht2016}.

\subsection{Strong-coupling optical effects of dark excitons and hot electrons}

\begin{figure}[ht!]
\centering
\includegraphics[width=1\linewidth,height=\textheight,keepaspectratio]{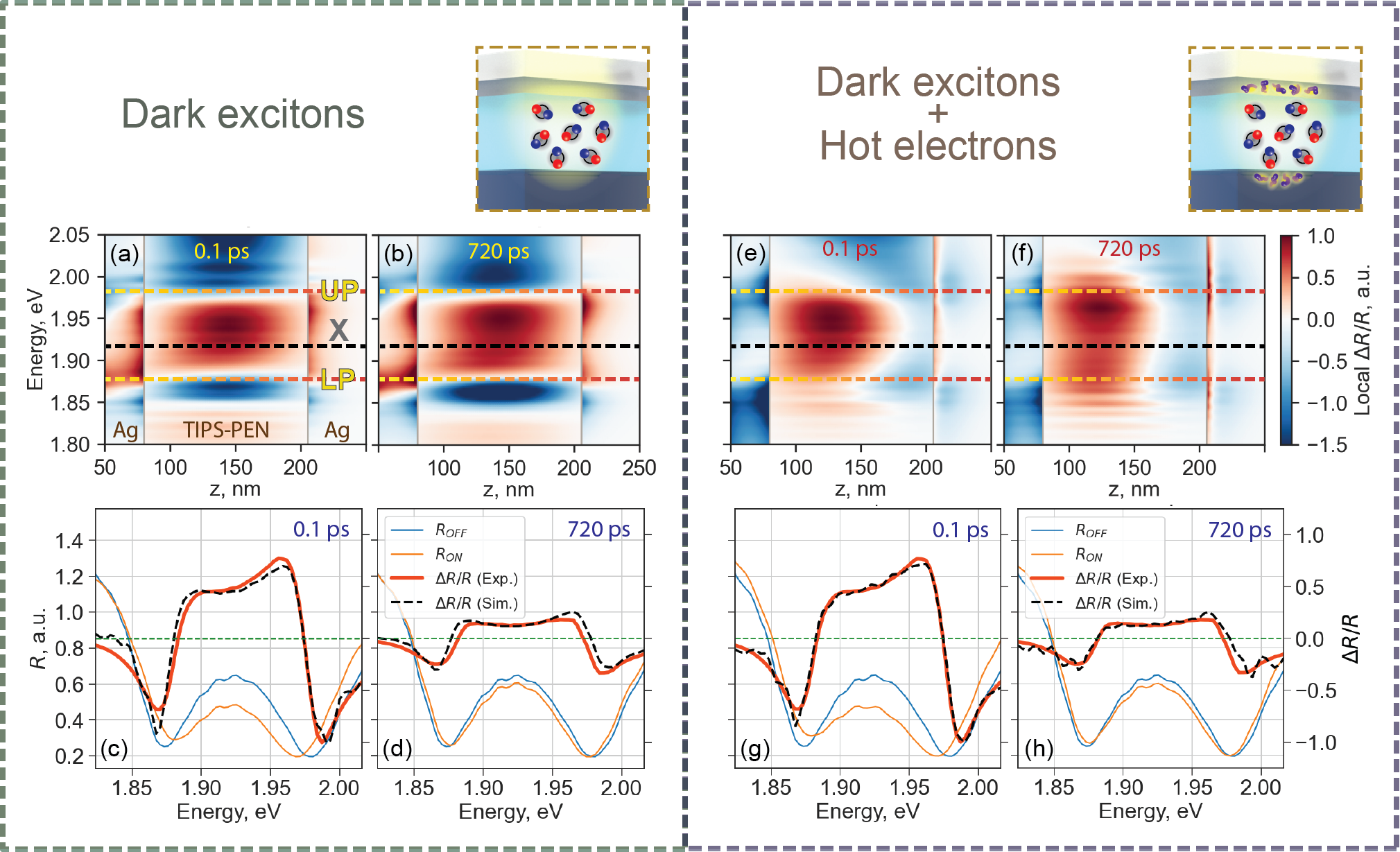}
\caption{Optical effects of dark excitons and free electrons. Simulated and normalized axially-resolved (a,b,e,f) and axially-integrated (c,d,g,h) $\Delta{R}/R$ spectra for (a,c,e,g) 0.1 and (b,d,f,h) 720~ps pump-probe delays. In (a), metal (Ag) and organic (TIPS-PEN) layers as well as UP, LP and dark-exciton peak $X$ (dashes) are indicated. (c,d,g,h) Fittings (black dashes) of experimental data (red solid lines) together with the extracted $R_{ON}$ (orange) and $R_{OFF}$ (blue) spectra.}
\label{fig:fig3}
\end{figure}

The sole effects of dark excitons responsible for the nearly mirrored dynamics of the introduced phases are investigated first. Several fundamental refractive-index changes (Supplementary material, Section~S13,S14) are taken into account in the simulations allowing us to unambiguously identify that it is only changes of dark-exciton population that are significant for introducing opposite phases to UP and LP. The resulting dispersive patterns in the numerical axially-resolved DR spectra (Fig.~\ref{fig:fig3}a,b) are evident in the organic layer, extending into the metal films in accordance with Maxwell's equations. The splitting between UP and LP increases with pump-probe delay (see retrieved $R_{\textrm{ON}}$ spectra) due to more excitons available for coupling to (probe) photons. This effect, referred to as polariton contraction, was observed previously for vibration-polaritons\cite{Dunkelberger2018,Xiang2019}. The measured DR spectra were fitted with axially-integrated lineshapes (Fig.~\ref{fig:fig3}c,d) featuring very good fitting quality, although it is noticeably poorer compared to Fano- and Hamiltonian-based results indicating, as expected, that additional effects from metal layers must be taken into account. Nevertheless, since it was possible to fit reasonably well the measured response at multiple pump-probe delays with parameters associated solely with dark excitons, we conclude that the role of dark excitons is significant in shaping the polaritonic response. 

Next, in addition to changes in the organic-layer's refractive index, we account for optically induced perturbations in the metal films (see Supplementary material, Section~S15, for more details). Fig.~\ref{fig:fig3}e,f show axially-resolved DR spectra at 0.1 ps and 720 ps delay. Within the organic layer, the general trend of increasing splitting between the polaritonic gain components with pump-probe delays can still be discerned. This indicates that dark excitons contribute as before, although to a lesser extent, since the effects of the metal layers also have to be accommodated. Reflectance loss in this case predominantly originates from the metal layers, especially, from the partially-transmitting 30-nm silver film ultimately, giving rise to a differential dispersive response (with narrower loss peaks, Fig.~\ref{fig:fig3}g,h). Retrieved reflectance and DR spectra are similar to those in Fig.~\ref{fig:fig3}c,d but yield significantly improved fitting quality comparable to the quality of Fano- and Hamiltonian-based fittings. To check for the sensibility of the retrieved reflectance spectra $R_{\textrm{ON}}$ and $R_{\textrm{OFF}}$, we experimentally confirmed their behaviour at short and long pump-probe delays using similar cavities (see Supplementary material, Section~S16).

The revealed strong absorptive contribution from metal films is sensible, because metals exhibit orders of magnitude stronger nonlinearities compared to other materials, which are accessible in thin metal films coupled to Lorentz-like oscillators\cite{Owens2010} (polaritons in our case). This is because thin films transmit only a small fraction of light due to the limited penetration depth of the electric field into metals\cite{Becker1997,Black1999,Ma2007}, with the transmitted light carrying information on these nonlinearities. It could be argued that a small penetration depth would contribute only a small signal to the axially-integrated response; however, this is offset by the metal's very large absorption coefficient, which enhances the absorptive contribution.

It is now possible to qualitatively relate made observations (Fig.~\ref{fig:fig3}e,f) to the phases $\widetilde{\phi}_{\pm}$ extracted from the non-Hermitian description of the polaritonic response. These phases correspond to net quantities containing the response of both the metal ($\widetilde{\phi}_{Ag,\pm}\sim90\degree$) and organic ($\widetilde{\phi}_{\textrm{TIPS-PEN},-}\sim-45\degree$ and $\widetilde{\phi}_{\textrm{TIPS-PEN},+}\sim-135\degree$) layers (see Fig.~\ref{fig:fig2}a). The organic layer contributes mostly gain, whereas metals contribute mostly loss. We note that the relaxation of hot electrons can contribute transient phase offsets between the $R_{\textrm{ON}}$ and $R_{\textrm{OFF}}$ signals through the changing penetration depth of metal mirrors\cite{Metzner2020}, affecting the cavity optical length, position of the cavity mode, as well as its bandwidth\cite{Becker1997}. In this case, the two mirrors would also generally contribute different phase offsets\cite{Ma2007}, and it is not straightforward to disentangle and quantify phase contributions from the two metal films at this stage. Further investigations are needed.

Lastly, the large noise in the TM-retrieved dynamics (Supplementary material, Fig.~S31) prevented us from the unambiguous determination of hot electrons evolving in accordance with a two-temperature model\cite{Temnov2009,Liu2021} where the majority of the dynamics occurs within the first time constant ($\sim$1~ps, electron-electron scattering) and the rest of the dynamics occurs within the second time-constant ($\sim$10~ps, electron-phonon scattering). To further support the presence of hot electrons, we performed a more powerful statistical correlation analysis of a large pump-probe dataset, which is described next.

\subsection{An imprint of hot-electron dynamics in the polaritonic response}

\begin{figure}[ht!]
\centering
\includegraphics[width=1\linewidth,height=\textheight,keepaspectratio]{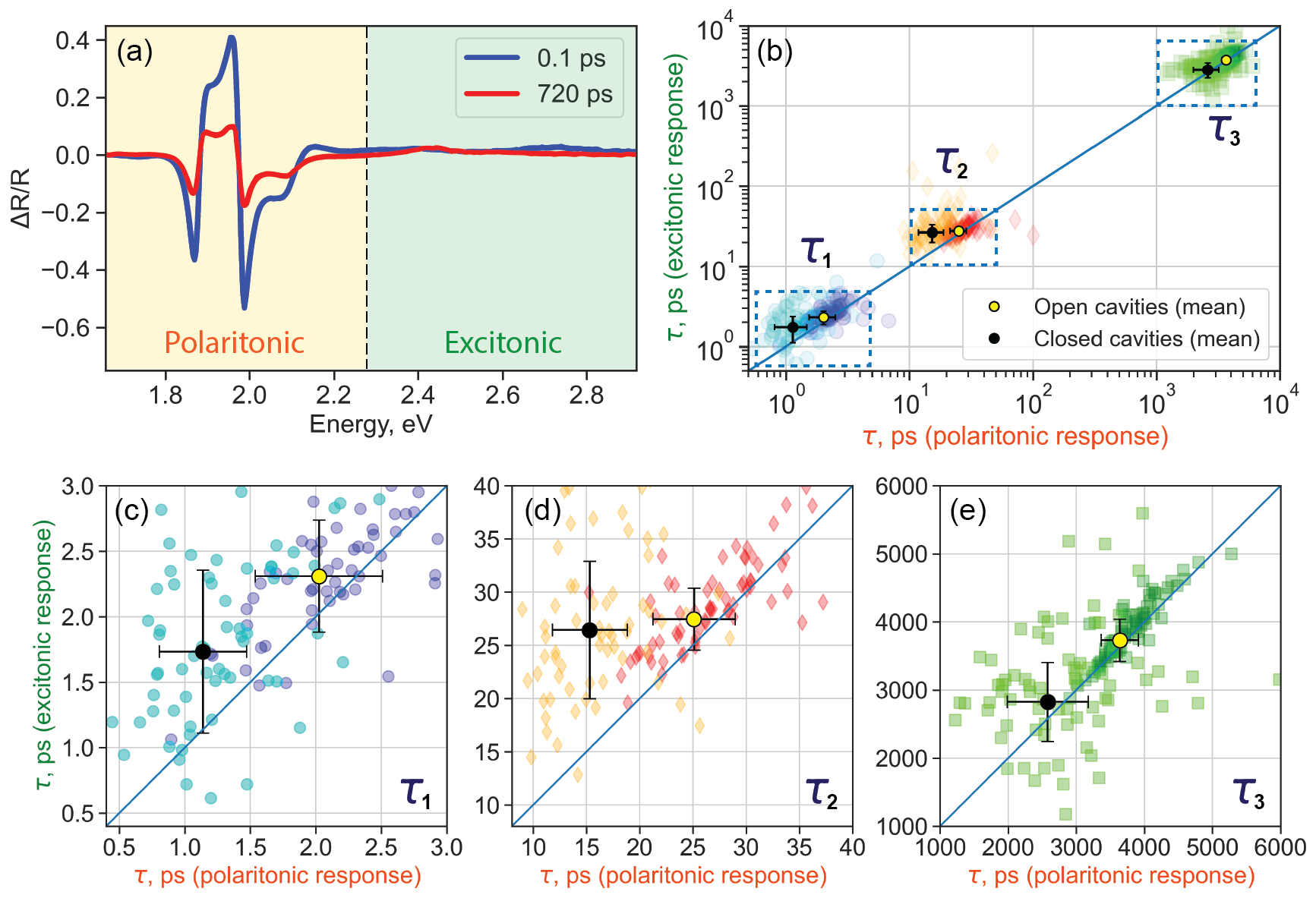}
\caption{A footprint of hot electrons. (a) Separation of transient spectra into polaritonic (shaded orange) and molecular (shaded green) response. (b--e) Correlation of time-constants extracted from molecular and polaritonic responses.}
\label{fig:fig5}
\end{figure}

In order to obviate the presence of hot-electrons' two-temperature dynamics in the polaritonic response, we conducted additional measurements of a large pump-probe dataset under various experimental conditions (see Supplementary material, Section S17, for more details). Each dataset was divided into two segments (Fig.~\ref{fig:fig5}a): a high-energy section containing a weak excitonic response and a lower-energy section containing a strong polaritonic response. Time constants were then extracted from these segments through a global analysis of pump-probe spectra\cite{Stokkum2004, Kolesnichenko} and correlated separately for excitonic and polaritonic responses in both open and closed cavities (Fig.~\ref{fig:fig5}b). Measurements on open OMs were used as a statistical reference.

In open cavities, as expected, the average time constants (yellow markers) primarily fell along the diagonal, as both lower- and higher-energy responses corresponded to mere exciton dynamics with no polaritonic effects were at play. However, in closed cavities, all three time constants (black markers) became shorter, with the first two significantly deviating from the diagonal. The shorter time-constants in the closed cavities can result from two effects: (\textit{i}) the Purcell effect\cite{Kolesnichenko}, which alone would yield time-constants $\tau_{Purcell,i}$ $i=1,2,3$, and (\textit{ii}) hot electrons relaxing in accordance with a two-temperature model (with time-constants $\tau_{Ag,i}$ $i=1,2$). The contribution of the latter effect is indeed plausible since the first two time-constants ($\tau_i,i=1,2$) measured in experiments are of the same order of magnitude as typical time-constants describing the dynamics of hot electrons in metals ($\tau_{Ag,1}\sim1$~ps, $\tau_{Ag,2}\sim10$~ps). Therefore, the measured time-constants $\tau_i$ ($i=1,2$) reflect net values such that $\min{(\tau_{Ag,i},\tau_{Purcell,i})}<{\tau}_i<\max{(\tau_{Ag,i},\tau_{Purcell,i})}$. However, it is possible to disentangle the two effects since the Purcell effect is characterized by a time-independent local density of photonic states but not the real number of photons in the cavity. Therefore, the shortening of the three time-constants should be by a similar (if not the same) factor, whereas the effect of hot electrons is time-dependent. Indeed, Fig.~\ref{fig:fig5}c--e show that the three time-constants are displaced horizontally by different amounts, with the first two being displaced to a greater extent than the third time-constant confirming the contribution from hot electrons. With all insights gained, the observed differences in the loss signal before the 10~ps pump-probe delay for thin and thick cavities (Supplementary material, Figure~S1e,f) are attributed to the dynamics of hot electrons in different photonic environments modulating the absorptive response in the metal films.

\section{Conclusion}

In summary, we investigated the polaritonic response from low-Q organic optical microcavities with TIPS-PEN and silver layers using steady-state and ultrafast-spectroscopy experiments. Theoretical analysis of the polaritonic gain-loss spectra indicated that dark-exciton fission governs the dynamics of the Rabi splitting by introducing opposite phases to upper and lower polaritons. Additionally, polaritons delocalized over both organic and metal films are shown to be sensitive to hot electrons evolving in accordance with a two-temperature model. These contributions resulted in doubly-dispersive Fano-like spectra, reflecting an intricate interplay between metal and organic films.

The approach outlined in this study holds promise for endeavors to distinguish polaritons and other cavity excitations in a variety of other microcavity structures. Our research outcomes carry significance for attempts to improve designs of optical microcavities to efficiently suppress or harvest such excitations, benefiting various fundamental and practical applications, such as polariton-electron and plasmon-electron coupling phenomena, as well as photonic manipulation of photophysical and photochemical processes. This is vital for the potential utilization of microcavities in novel devices such as energy-efficient organic solar cells.

Two more specific applications are envisioned. Firstly, the optical messengers of dark-exciton fission and polaritonic response are energetically far from each other. The observed fission sensitivity at polaritonic energies indicates the possibility of translating characteristic spectral window of physical phenomena that are challenging to detect to other energies, simply by tuning the cavity resonance to overlap with correlated transitions that are more convenient for detection. This could be exploited in various sensing applications. Secondly, the sub-ns polaritonic response represents the mathematical operation of differentiation of polaritonic spectra. This could be viewed as an alternative to analogue  differentiation using ultrashort optical pulses, which could ultimately be exploited, \textit{e.g.}, in polariton-based spatial light modulators\cite{JingQuan2008} for spectral differentiation of input beams as part of derivative-based analyses\cite{ZangenehNejad2020,dubrovkin2021}.

\section{Methods}

\subsection{Sample preparation}

Fabry-Pérot cavities were prepared on glass substrates (AF32eco, Schott, 25x20x0.5 mm$^3$) cleaned by sonication (10 minutes) in alkaline solution (Hellmanex), deionized water and 2-propanol. 100-nm silver films were deposited by thermal evaporation. TIPS-PEN (6,13-Bis(triisopropylsilylethynyl)pentacene, Ossila Ltd.) and polystyrene (PS, Mw ca. 222k, Polymer Source Inc.) were used as received. In manufacturing the active layer, PS was dissolved in toluene at a concentration of 20 mg.mL$^{-1}$, and TIPS-PEN was added to the solution with relative mass ratio of 30\%. Thin films of TIPS-PEN/PS were prepared by spincoating and their thicknesses were adjusted with the spin rate (1800 and 2000 rpm for the thick and thin sample, respectively). The organic layer thicknesses of thin and thick cavities are 139~nm and 148~nm, respectively. The thin films were then annealed for 1 minute at 100$\degree$C to remove traces of residual solvent. Then, 30~nm silver films were thermally evaporated on top. Finally, 50~nm protective layer of aluminium oxide (AlO$_x$) were deposited on top of the cavities via atomic layer deposition (Ultratech, Savannah S100, precursor trimethyl-aluminium, Strem Chemicals, Inc.) at 80$\degree$C.

\subsection{Angle-resolved steady-state reflectance spectroscopy}

Absorption spectrum of TIPS-PEN/PS thin film was acquired with an Agilent Cary 6000i UV-Vis-NIR absorption spectrometer.
Angle-resolved steady-state reflectance spectra of the Fabry-Pérot cavities were acquired in a home-built Fourier microscopy setup\cite{Graf2016} using a halogen light source (Ocean Optics, HL-2000-HP) focused onto the sample via an objective (60x, NA=0.9, Olympus, MPLAPON60X). The reflected light was imaged from the back focal plane of the objective onto the entrance slit of a spectrometer (Princeton Instruments IsoPlane SCT 320). A linear polarizer was used to select TE polarization. The resulting spectra were acquired with a Si-CCD camera (Princeton Instruments, PIXIS:400).

\subsection{Transient reflectance spectroscopy}

Transient reflectance spectroscopy performed in this work is described elsewhere\cite{Kolesnichenko}. In a nutshell, an amplified Ti:Sapphire oscillator (Coherent Libra) produced 80~fs 1.55~eV optical pulses at 1~kHz. These in turn were split to drive a lab-built noncollinear optical parametric amplifier generating pump pulses (39~fs, 1.91~eV, 20~nJ, p-polarized), and CaF$_2$ white-light generator generating p-polarized probe pulses covering 1.65--2.88~eV. Differential reflectance
\begin{equation}
\begin{aligned}
\frac{\Delta{R}}{R} = -\frac{R_\textrm{ON}(E,\tau)-R_\textrm{OFF}(E)}{R_\textrm{OFF}(E)}
\end{aligned}
\label{eqn:DR}
\end{equation}  
was acquired, where $R_\textrm{OFF}$ is reflectance of the probe beam from unperturbed sample, $R_\textrm{ON}$ is that from the pumped sample, $\tau$ is pump-probe delay, $E$ is detection energy. The pump and probe beams were incident on the samples at angles 4$\degree$ and 1$\degree$, respectively. The resolution of the experiment was 100 fs.

\medskip
\bibliography{main.bib}

\medskip
\subsection{Acknowledgements} \par 
This research was funded by the Deutsche Forschungsgemeinschaft (DFG) via the Sonderforschungsbereich SFB 1249-B04 and SFB 1249-C09.

\subsection{Author contributions}
M.H. and J.Z. provided samples; M.H. conducted steady-state spectroscopy experiments, analyzed and interpreted results; P.V.K. conducted ultrafast spectroscopy experiments, analyzed and interpreted results; F.H. assisted with data analysis; P.V.K., M.H., O.K. and T.B. participated in discussions; P.V.K. and T.B. wrote the manuscript;  P.V.K., M.H. and T.B. revised the manuscript; T.B. secured funding.

\subsection{Authors disclosure statement} \par 
The authors declare no competing interests.

\subsection{Figure captions}

\textbf{Figure~\ref{fig:fig1}}. Steady-state and transient spectra of thin microcavity. (a) Normalized angle-resolved steady-state reflectance spectrum, with the upper and lower polaritons (UP and LP), dark excitons ($X$ and $X'$), and cavity mode (\textit{C}) indicated. Dashed lines represent uncoupled excitonic (horizontal lines) and photonic (parabola) states; solid yellow lines represent hybridized (polaritonic) states. (c) Steady-state absorption spectrum of the cavity and bare TIPS-PEN (dashed and shaded in grey), and pump spectrum (shaded in yellow). (b) Transient reflectance experiment. The inset (right) shows dark excitons in the organic layer and hot electrons in metal films "dressed" by polaritonic field. (d) Transient reflectance spectrum at short (0.1~ps, blue) and long (720~ps, red) pump-probe delays. The weak molecular response (dashed) is scaled by a factor of 10.
\\
\\
\textbf{Figure~\ref{fig:fig2}}. Non-Hermitian Hamiltonian description of transient reflectance spectra. (a) A cartoon explaining the physical meaning of the dynamic phases $\phi_{\pm}$ after remapping them onto more intuitive values $\widetilde{\phi}_{\pm}=90\degree+2\phi_{\pm}$ (modulo 180$\degree$). (upper row) "Pump-ON" (red) and "pump-OFF" (blue) reflectance spectra. The inset shows the phase $\widetilde{\phi}$ and the direction of the displacement (red line) of the "pump-ON" spectrum relative to the "pump-OFF" spectrum. (b) Fittings (dashed lines) of short- (red) and long-delay (blue) spectra (solid lines).  (c--f) The scaling amplitude $A$ (c, solid), coupling strengths $V$ (f, solid), and phases $\widetilde{\phi}_{\pm}$ (e,f) retrieved from fittings for thin (dark lines) and thick (yellow lines) cavity. In (c) dashed lines represent the absolute value of the bleach signal from the corresponding open cavities (\textit{i.e.}, cavities without 30-nm metal film). In (f) dashed lines represent $T_1$~ESA in thin and thick cavity.
\\
\\
\textbf{Figure~\ref{fig:fig3}}. Optical effects of dark excitons and free electrons. Simulated and normalized axially-resolved (a,b,e,f) and axially-integrated (c,d,g,h) $\Delta{R}/R$ spectra for (a,c,e,g) 0.1 and (b,d,f,h) 720~ps pump-probe delays. In (a), metal (Ag) and organic (TIPS-PEN) layers as well as UP, LP and dark-exciton peak $X$ (dashes) are indicated. (c,d,g,h) Fittings (black dashes) of experimental data (red solid lines) together with the extracted $R_{ON}$ (orange) and $R_{OFF}$ (blue) spectra.
\\
\\
\textbf{Figure~\ref{fig:fig5}}. A footprint of hot electrons. (a) Separation of transient spectra into polaritonic (shaded orange) and molecular (shaded green) response. (b--e) Correlation of time-constants extracted from molecular and polaritonic responses.

\begin{suppinfo}

The following file is available as Supplementary material: 

Spectral characterization of thin and thick cavities; Estimation of the organic-layer thickness and refractive index; Hopfield coefficients; Local electric fields; Transient differential reflectance spectra; SVD decomposition of pump-probe spectra; Fano resonances: a phenomenological description of transient polaritonic response; More on non-Hermitian Hamiltonian formalism; Reduction of 3D Hilbert space to two dimensions; Non-Hermitian description of polaritonic response; Non-Hermitian description of polaritonic response with added dark-exciton bleach; Global exponential fitting of the parameters; Pump-induced refractive index of the organic layer; Optical simulations of the effects of dark excitons; Optical simulations of the effects of dark excitons and hot electrons; Experimental $R_{\textrm{OFF}}$ and $R_{\textrm{ON}}$ spectra; A comment on statistical correlation analysis of multiple pump-probe data.

\end{suppinfo}

\end{document}